\documentstyle[preprint,aps]{revtex}
\draft
\begin{document}
\title{Adiabatic quantum computation with Cooper pairs} 
\author{D.V. Averin}
\address{Department of Physics, SUNY at Stony Brook, Stony Brook, 
NY 11794}

\maketitle

\begin{abstract}
We propose a new variant of the controlled-NOT quantum logic gate 
based on adiabatic level-crossing dynamics of the q-bits. The gate 
has a natural implementation in terms of the Cooper pair transport 
in arrays of small Josephson tunnel junctions. An important advantage 
of the adiabatic approach is that the gate dynamics is insensitive 
to the unavoidable spread of junction parameters. 
\end{abstract}

\vspace{3ex} 

\narrowtext

An invention of quantum algorithms (for a review see, e.g., 
\cite{b1}) changed the foundations of the theoretical computer 
science by demonstrating that the information is processed 
differently by quantum and classical systems. In an ideal 
world, a quantum algorithm implemented on a quantum computer can 
radically outperform the classical algorithm by making use of 
quantum parallelism inherent in entangled quantum states. Examples 
of problems which can be efficiently solved with quantum algorithms 
include factorization of large numbers \cite{b2} and database 
search \cite{b3}. 

Practical realization of a quantum computer requires, however, 
very precise and reversible time evolution of complex quantum 
mechanical systems, the fact that gives rise to serious doubts 
\cite{b4} as to whether even the simplest version of a quantum 
computer will ever become a reality. It is therefore important 
to look into various possible ways of implementing simple 
elements of quantum computer -- quantum logic gates in order 
to find the optimal approach to building such a computer. 
Generally speaking, a quantum gate should satisfy two 
contradictory requirements: being isolated from the outside 
world in order to maintain quantum coherence, and interacting 
with other q-bits, read-out system, etc., in order to perform 
a meaningful computation. Existing quantum gate proposals use 
various systems including ion traps \cite{b7,b8}, electrodynamic 
cavities \cite{bd1}, semiconductor quantum dots \cite{bd2,bd3}, 
NMR spectroscopy \cite{bd4}, quantum flux dynamics in SQUIDs 
\cite{bdd}. Some of these proposals, for instance, ion-trap or 
NMR, are characterized by potentially very long relaxation times, 
since the gates in these proposals are well isolated physically 
from the outside world. However, due to the very same reason, 
these gates can not be combined easily to form larger computing 
systems. For other gates, for instance, based on semiconductor 
quantum dots, the situation is the opposite. In principle, it is 
not too difficult to integrate them into larger structures, but 
there seems to be very little hope of reducing decoherence rates 
to a level acceptable for quantum computation.     
 
The aim of this work is to suggest another possible realization of 
quantum gates which is based on manipulation of the charge states of 
small Josephson tunnel junctions. This approach combines both the 
potential for relatively long relaxation times and large degree of 
design flexibility, and probably represents one of the best, if not 
the best, hope for realization of a quantum computer of medium 
complexity. Such a computer, while not being sufficient for 
factorization of large numbers of practical interest, could be 
sufficiently complex to perform privacy amplification \cite{b5,b6} 
in quantum communication channels.  

The basic universal set of quantum logic gates consists of 
the one q-bit gates and two q-bit controlled-NOT (CN) gate. 
In practically any implementation, including the one discussed 
below, the dynamics of the two q-bit gates contains all elements 
of the one q-bit operation, and therefore, we can concentrate on 
the two q-bit CN gate.  The operation of this gate can be 
described simply as inversion of the target q-bit states 
when the control q-bit is the ``1'' state. The state of the control 
q-bit should be unchanged during this operation. In the standard 
approach, the CN-operation is achieved through the use of the 
ac-driven Rabi transition between the q-bit states \cite{b7,b8,bd4}. 
We propose another general scheme of the CN gate which uses 
adiabatic transitions between the q-bit states and is  
more suitable for implementation in systems of small tunnel 
Josephson junctions. 

The main idea of the adiabatic CN-gate is as follows. 
Interaction between the control and target q-bit makes 
the energy difference between the two states of the target q-bit 
dependent on the state of the control q-bit. If the control q-bit 
is in the state ``1'' of the computational basis, the energy
difference is smaller and under application of the time-dependent 
bias the target q-bit passes through the level-crossing point,  
where the energies of its two states are equal -- see Fig.\ 1. 
If the rate of the bias increase is sufficiently small, the 
two states of the target q-bit exchange their occupation 
probabilities. When the control q-bit is in the ``0'' state, the 
energy difference is larger and the same bias pulse does not   
drive the target q-bit through the level-crossing point. In this 
case, 
the occupation probabilities remain the same. The tunnel coupling 
between the states of the control q-bit is suppressed during the 
entire process so that their occupation probabilities do not 
change in either case. This time evolution realizes CN-gate 
operation provided that the parameters of the two q-bits are 
chosen in such a way that the dynamic phases accumulated in the 
system evolution along all four ``paths'' are equal. 

To implement this dynamics of the two q-bit system we need to 
control both the energy difference $\varepsilon_j(t)$ between the 
two states of a q-bit in the computational basis $|0\rangle$ and  
$|1\rangle$, and the tunneling amplitude $\Omega_j(t)$ in this 
basis, i.e., the Hamiltonian of the system should be: 
\begin{equation} 
H=\sum_{j=1,2} (\varepsilon_j(t)\sigma_{zj} +\Omega_j(t)
\sigma_{xj}) + \eta \sigma_{z1}\sigma_{z2} \, , 
\label{1} \end{equation} 
where $\sigma_j$ are the Pauli matrices for the $j$th q-bit, 
and $\eta$ is the energy of interaction between the q-bits.  
Although the basic gate dynamics does not require 
the modulation of the interaction strength $\eta$, this interaction 
does not allow to make the energies of all four gate states 
equal after the gate operation. This means that the 
relative phases of these states will continue to evolve at a rate 
on the order of $\eta/\hbar$ and we need to be able to manipulate 
the gate on a time scale much shorter than $\hbar/\eta$, which in 
its turn should be much shorter than the time scale of the adiabatic 
dynamics. This long hierarchy of time scales presents a serious 
problem that exists for other proposals of quantum gates as well. 
A better solution is to design a gate in a way that allows to 
switch the interaction on and off, despite the fact that this makes 
the design appreciably more complicated.  

If the interaction energy $\eta$ in the Hamiltonian (\ref{1}) can be 
controlled, we can separate the gate dynamics into three 
steps. At first, the two q-bits are brought into contact by switching 
on $\eta$ and $\Omega_2$ ($\Omega_1$ is completely suppressed 
throughout the gate operation). Simultaneously the energy difference 
$\varepsilon_2$ between the states of the target q-bit is set to some 
nonvanishing initial value. Then this energy difference is increased 
while all other energies are kept constant. This step is the central 
``level-crossing'' part of the gate dynamics. During the final third 
step both $\eta$ and $\Omega_2$ are suppressed back to zero so that 
the two q-bits are effectively separated and $\varepsilon_2$ can  
also be reduced to zero. 

The precise functional dependence of $\Omega_2$, $\varepsilon_2$, 
and $\eta$ on time does not qualitatively affect the gate dynamics, 
as long as all these parameters are changed gradually. The limitation 
on the rate of the parameter variations is associated with the 
unwanted transitions between the instantaneous energy eigenstates of 
the system which are brought about by these variations. 
These transitions violate the correct adiabatic dynamics which 
assumes that the system remains at all times in the same eigenstate 
it occupied initially. Adopting a simple model time dependence of 
the energy difference $\varepsilon_2(t)$:  
\begin{equation} 
\varepsilon_2 (t) = \varepsilon + u \tanh (t/\tau) \, , 
\label{2} \end{equation} 
and using the standard quasiclassical approach \cite{b9} we can  
calculate explicitly the probability $p$ that the system makes 
an unwanted transition during the central second step of the gate 
operation. This simple calculation confirms the expected result 
that the probability $p$ reaches maximum when the system passes 
through the level crossing-point and is given then by the standard 
Landau-Zener expression: 
\begin{equation} 
p_{LZ}= \exp \{ -\frac{\pi \tau \Omega^2}{\hbar u} \} \, . 
\label{4} \end{equation} 
Here $\Omega$ is the magnitude of the tunnel amplitude 
$\Omega_2(t)$ that is kept constant during this step of the gate 
operation. Thus, the condition $p_{LZ}\ll 1$, i.e.,  $\tau\gg 
\hbar u/\Omega^2$, ensures the correct adiabatic dynamics of the 
gate. 

This implies that dynamics of the occupation probabilities 
of the states of adiabatic CN gate is not sensitive to the precise 
values of the parameters in the Hamiltonian (\ref{1}) provided that 
they satisfy several constrains which ensure the gate operation 
shown in Fig.\ 1:  
\begin{equation} 
\eta+u-\varepsilon \gg \Omega \, , \;\;\;\; \eta -u -\varepsilon 
\ll - \Omega \, , \;\;\;\; u -\eta -\varepsilon \ll - \Omega \, . 
\label{5} \end{equation} 
If all these conditions are satisfied, the evolution of the 
absolute values of the occupation amplitudes $\alpha_{ij}$ of the 
four gate states corresponds to the correct CN operation:   
\[ |\alpha_{0j}| \rightarrow  |\alpha_{0j}| \, , \;\;\;  
|\alpha_{10}| \rightarrow  | \alpha_{11} |\, ,\;\;\; 
|\alpha_{11}| \rightarrow  | \alpha_{10} | \, . \]
(The indices $i$ and $j$ denote the states of the control and target 
q-bit respectively.)  
Besides this time evolution of the absolute values of $\alpha_{ij}$, 
the correct gate dynamics requires also that phases of the four 
states accumulated in the process of the gate evolution are equal 
modulo $2\pi$. This can be achieved by adjusting the bias 
$\varepsilon_{1,2}$ of the two q-bits and the energy splitting 
$\Omega_2$ during the gate operation. The bias $\varepsilon_1$ 
controls the relative phases of the pairs of states evolving 
from the 0 and 1 state of the control q-bit, while 
$\varepsilon_2$ and $\Omega_2$ control the phases within each 
pair. With such an adjustment of the phases, the adiabatic time 
evolution of the coupled q-bits represents correctly the CN 
quantum logic gate. 

This gate can be naturally implemented in systems of small 
Josephson tunnel junctions in the Coulomb blockade regime --       
see, e.g., \cite{b10,b11}. The energy diagram of an elementary 
building block of such a system, a single junction, is shown in 
Fig.\ 2. The dominant contribution to the junction energy is given 
by the charging energy $U(n)$ of the junction as a capacitor: 
\[ U(n) =  \frac{(2en-Q_0)^2}{2C} \, , \] 
where $C$ is the junction capacitance, $n$ is the number 
of Cooper pairs transferred across the junction, and $Q_0$ 
is the charge induced by the external bias voltage $V_0$ across 
the junction, $Q_0 = V_0C$. In general, the states with different 
$n$'s are separated by large energy gaps on the order of 
elementary charging energy $E_C= e^2/2C$. However, when 
the external voltage $V_0$ induces the charge of approximately 
one electron on the junction capacitance, $Q_0 \simeq e$, 
the two state, $n=0$ and $n=1$ are nearly degenerate and are 
separated from all other states by the large energy gaps -- see Fig.\ 
2. In this regime the junction behaves effectively as a two-level   
system. The energy difference $\varepsilon$ between the level of this 
two-level system is controlled by the external voltage $\varepsilon = 
2e(e-Q_0)/C$, while the amplitude $\Omega$ of tunneling between them 
is determined by the Josephson coupling energy $E_J$ of the junction, 
$\Omega=E_J/2$. The Josephson coupling energy depends on the tunnel 
resistance $R_T$ of the insulator barrier between the electrodes, and 
for the electrodes with equal superconducting energy gaps $\Delta$ is 
equal to $\pi \hbar \Delta/4e^2 R_T$ -- see, e.g, \cite{b11}.  

Thus, the appropriately biased small Josephson tunnel junction 
is a macroscopic two-level system, with the two states represented 
by the position of a single Cooper pair on the left or right 
electrode of the junction. In principle, this system can be used as 
a q-bit of the quantum logic gates. However, if q-bits are 
represented with single junctions, neither the tunneling 
amplitude $E_J/2$ nor the coupling strength of the two q-bits which 
is determined by the coupling capacitance between the junction 
electrodes can be modulated in time as required by the design of 
the adiabatic CN gate. In particular, to realize adiabatic 
dynamics it should be possible to suppress both the tunneling 
amplitude and interaction strength to zero between the active cycles 
of the gate operation. This problem can be circumvented if q-bits 
are represented not with individual junctions but with the 
one-dimensional arrays of junctions. In an array, the tunneling 
amplitude $\Omega$ between the two islands of the array can be 
effectively modulated by the gate voltages applied to the islands 
of the array, and the interaction energy $\eta$ of charges in the 
array decreases exponentially with the distance between them.

To make a q-bit out of a uniform array, all islands should have 
individual gate electrodes supplying the gate voltages $V_j$ (Fig.\ 
3a,b), and two internal islands of the array should be biased with 
the voltages 
$\pm e/C_t$, where $C_t=(C_0^2+4CC_0)^{1/2}$ is the total 
capacitance of an internal island in the array -- see, e.g., 
\cite{b10}, and $C$, $C_0$ are, respectively, the junction 
capacitance, and the capacitance between each island and its 
gate electrode (Fig.\ 3b). These voltages induce the charges $e$ 
and $-e$ on the two islands, so that the two charge configurations 
of the array: one with no Cooper pair transferred across any 
junction and another one with a Cooper pair transferred between the 
two biased islands, from $e$ to $-e$, have the same energy. This 
means that if the bias conditions do not deviate strongly from these 
conditions, all other charge configurations of the array have much 
larger energies and the array dynamics is equivalent to the 
two-state dynamics that can be described in terms of the tunneling 
of a single Cooper pair between the two islands. In this regime 
the array can be viewed as a q-bit with the two positions of the 
Cooper pair on one or another island representing the two states 
of the computational basis of this q-bit.   

If the two islands containing the q-bit states are separated by 
$m$ junctions, the amplitude of tunneling $\Omega$ between them 
depends exponentially on the separation $m$. The dominant 
contribution to $\Omega$ comes from the process in which the 
Cooper pair is transferred sequentially through the junctions 
separating the islands, and can be written as:   
\begin{equation}  
\Omega= \frac{E_J}{2}\prod_{k=1}^{m-1} \frac{E_J}{2E_k} \, ,
\label{6} \end{equation}  
where $E_k$ are the energies of the intermediate charge 
configurations resulting from the Cooper pair transfer through the 
first $k$ junctions. These energies are controlled by the gate 
voltages applied to the intermediate islands. 

The most important feature of the Cooper pair states forming 
q-bit basis is that they can be moved along the array by the 
adiabatic level-crossing transitions similar to those discussed 
above. A Cooper pair is transferred between the two adjacent 
islands when a gate voltage of the initially occupied island is 
increased/decreased while the gate voltage of the neighboring 
island is decreased/increased adiabatically past 
$e/C_t$. The adjacent islands are coupled by the tunneling 
amplitude $E_J/2$, and the Cooper pair is transferred with 
the probability exponentially approaching one if the rate of 
change of the 
gate voltages is small on the scale of this amplitude. Similar 
manipulation of the gate voltages also shifts the empty state 
of the q-bit by one island. In this way it is possible to move 
the q-bit states around, either shifting both states along the 
array, or changing the separation $m$ between the two states. 

This dynamics is analogous to the one used in the so-called 
single-electron \cite{b12} and single Cooper pair \cite{b12*} pump,  
or single-electron parametron \cite{b12**}. It allows to implement 
the general scheme of the 
adiabatic CN gate with the two coupled arrays representing the 
two q-bits of the gate (Fig.\ 3c). As a first step of the gate 
operation, the q-bit states in both arrays are moved towards the 
ends of the arrays where they can interact via the coupling 
capacitance $C_i$. The states of the controlled q-bit in the 
first array have sufficiently large separation $m$ so that 
their tunnel coupling $\Omega_1$ is negligible. By contrast, 
the states of the target q-bit in the second array are put on 
the adjacent islands in order to maximize their tunnel coupling, 
$\Omega_2 = E_J/2$. Then a pulse of the bias voltage is applied 
to the first junction of the target q-bit array. If the control 
q-bit is in the ``1'' state, a Cooper pair occupies the island 
of the first array closest to the second array and creates 
additional potential drop $\delta V$ across the junction of 
the target q-bit: 
\begin{equation} 
\delta V = \frac{8e C_i}{(C_0+C_t+2C)(C_0+C_t+4C_i)} \, . 
\label{8} \end{equation}  
In this case the bias pulse drives the target q-bit through 
the level-crossing point so that the occupation probabilities of 
its states are interchanged. When the control q-bit is in the 
``0'' state, the Cooper pair of this q-bit is inside the array and 
does not produce extra voltage across the target q-bit junction,  
which then does not reach the level-crossing point, and the 
occupation probabilities of its states remain the same. 
During the last step of the gate operation it 
is returned to its initial configuration, i.e., the separation 
of the states of the target q-bit is increased to suppress the 
tunnel amplitude $\Omega_2$ to zero, and the states of the 
both q-bits are shifted inside the arrays. Then the interaction of 
the q-bit states becomes negligible due to screening by the gate 
electrodes, which is known to lead to the 
exponential suppression of the interaction energy $\eta$ between 
two Cooper pairs separated by $m$ junctions of one array 
\cite{b10}: 
\begin{equation}
\eta = \frac{(2e)^2}{C_t} \lambda^m\, , \;\;\;\; \lambda = 
\frac{2C}{2C+C_0+C_t} \, . 
\label{7} \end{equation} 

This implementation of the CN quantum gate can only be practical 
if it is stable against deviations of the real gate structure from 
the idealized model used above. Such deviations are fundamentally 
unavoidable in all macroscopic realizations of quantum gates. 
For instance, the real electrostatics of the Josephson junction 
gate is much more complicated that the model characterized by the 
two nearest--neighbor capacitances $C$ and $C_0$. It involves full 
capacitance matrix $C_{ij}$ in which even remote islands interact 
with each other, and should also describe small fluctuations of the 
nearest-neighbor capacitances around their average values. 
An important advantage of the adiabatic approach  
is that these complications can be compensated for by the adjustment 
of the bias voltages and do not change qualitatively the gate 
dynamics. Indeed, the adiabatic transfer of a Cooper pair depends 
only on the resonance condition that the energies of all Cooper 
pair states along the array are the same, which ensures correct 
transfer of the occupation probabilities of the gate states. 
The bias voltages can always be tuned to satisfy the resonance 
condition regardless of the form of the capacitance matrix. A 
practical proof of this statement is provided by the experimentally 
demonstrated operation of a similar system, single-electron pump, 
with accuracy better than $10^{-6}$ \cite{b19}.  

The only instance when the gate dynamic relies heavily on the 
simplified model of the array electrostatic is in the 
assumption of the exponential screening of the electrostatic 
interaction inside the array. In the realistic model 
of electrostatics, interaction at large distances depends on the 
external environment of the array. The exponential screening of 
the interaction can still be obtained even in this case, but 
requires that the array is placed between the two conducting 
ground planes. 
 
These considerations show that dynamics of the occupation 
probabilities of the gate states is indeed insensitive to the week 
disorder in the gate parameters. However, the proper dynamics 
of the system as quantum logic gate requires also that the phases 
of the occupation amplitudes accumulated during the gate operation 
are all equal modulo $2\pi$. In this respect, fluctuations of the 
junction parameters do present a problem since they make the dynamic 
phases of the gate states unpredictable. This problem can be resolved 
if the disorder in the parameters is static on a sufficiently 
long time scale. In this case, the phases can be measured and 
compensated for by the fine-tuning of the gate voltages. 

In order to measure the phases, we need to transform them into the 
occupation probabilities of the gate states which in their turn can 
be measured with a singe-electron electrometer (see, e.g., 
\cite{b10*}, Chapter 9). An electrometer measures an average charge 
of the island and therefore gives information about the occupation 
probabilities of the gate state, but is insensitive to the phase of 
the occupation amplitudes. 
Suppose that as a result of a prior measurements, we know that 
the occupation probabilities of the two q-bit states are $p_1$ and 
$p_2$ . The two states are decoupled (the corresponding tunneling 
amplitude $\Omega$ is zero) and their energies are equal, so that 
there is some stationary phase difference $\varphi$ between their 
occupation amplitudes. The phase $\varphi$ can be transformed into 
the occupation probability by rotation $\hat{U}$ of the q-bit states 
in the Hilbert state, $\hat{U} =\exp \{ i \pi \sigma_x/4 \}$. 
This rotation is achieved if the barrier between the states is 
reduced temporarily in such a way that 
\[ \int dt \Omega(t) =\pi \hbar /4 \, . \]  
The resulting occupation probabilities 
\[ q_{1,2} = 1/2 \mp (p_1 p_2)^{1/2} \sin \varphi \, , \]  
depend on the phase $\varphi$, and by measuring them we can measure
 $\varphi$. After the phase is known it can be compensated for by 
adding an extra voltage pulses at the end of the gate operation. 
With this fine-tuning, the gate dynamics becomes effectively 
independent of the static disorder in the gate parameters. 

The above discussion assumes that the energy relaxation and 
associated with it time-dependent fluctuations of the phase are 
negligible. There are several dissipation and dephasing mechanisms 
in the Josephson tunnel junction systems. Some of them are well 
understood and can be controlled within certain limits. One of 
these mechanisms is the quasiparticle tunneling. In general, it 
coexists with the Cooper pair tunneling and makes junction 
dynamics irreversible. However, if both the temperature $T$ and 
charging energy $E_C$ of the junctions are much smaller than the 
superconducting energy gap $\Delta$, the quasiparticle tunneling 
is suppressed by the parity effects \cite{b13,b13*,b13**} to a 
level where it can be negligible on the macroscopic time scales 
\cite{b14,b15}. Another dissipation mechanism is coupling to the 
electromagnetic excitations supported by the system of 
superconducting electrodes. A Cooper pair oscillating between the 
two islands creates oscillating currents in the islands and 
electric fields around the islands which couple to these modes. 
The power $P$ lost to electromagnetic modes depends on the 
specific geometry of the islands and connecting them tunnel 
junctions. Part of the losses comes from the direct dipole 
radiation from the junctions and can be estimated as radiation of 
a dipole of length equal to the length $d$ of the junction 
electrodes: 
\begin{equation} 
P_{d} \simeq \frac{e^2 \omega^4 d^2}{4\pi \epsilon_0 c^3} \, . 
\label{9*} \end{equation}  
The radiated power is not exponentially small, nevertheless it  
decreases sufficiently rapidly with decreasing ratio of the island 
size to the radiation wavelength $\lambda \simeq c/\omega$ at 
frequency $\omega \simeq E_J/\hbar$. Therefore, to keep this type 
of radiation losses small the islands of the junction arrays should 
be much smaller than the wavelength at frequency $E_J/\hbar$, 
the condition that is always satisfied in small junctions.  

The crucial contribution to radiation losses comes from the coupling 
to electromagnetic modes supported by essentially ``infinite'' 
external gate electrodes supplying bias voltages to the islands. 
In the relevant regime with $C_0\ll C$, the power dissipated into 
these modes can be estimated in terms of the wave impedance $\rho$ 
of the gate electrodes as   
\begin{equation} 
P_{l} \simeq (\frac {eC_0}{C})^2 \omega^2 \rho \, . 
\label{9**} \end{equation}  
We see that this dissipation mechanism limits the magnitude of the 
island capacitance to the gate electrodes $C_0$. In the simple model 
of the gate electrostatics, $C_0$ determines also the number of 
islands of the junction array that are polarized by a single Cooper 
pair, and restriction on $C_0$ translates into the limitation on how 
small the number of junctions in the arrays can be. If however, one 
introduces ground planes which give rise to extra stray capacitances 
of the array islands, these two limitations becomes uncoupled. In any 
case, for realistic values of the parameters (see the estimates below) 
the losses (\ref{9**}) in the external electrodes should give the 
dominant contribution to decoherence for the Cooper pair tunneling.  

Besides these ``controllable'' mechanisms of dissipation that 
depend on the gate geometry, the Cooper pair tunneling in the 
junction arrays is affected also by the ``internal'' dissipation in 
all elements of the arrays. The most important source of noise and 
dissipation of this kind is the $1/f$ charge noise in the insulators 
surrounding the junctions: substrate and tunnel barriers. The 
strength of the noise is material dependent and can not be  
estimated from first principles. Experimentally, characteristic 
time scale of the charge noise varies from millisecond range 
\cite{b16} to seconds and hours \cite{b17}, and is much longer that 
characteristic time of the Cooper pair tunneling $\hbar/E_J$ which 
determines the rate of the gate operation. Therefore, the gate 
can go through the large number of cycles of operation before the 
decoherence due to the charge noise starts to affect its dynamics. 

Before concluding, we summarize the conditions that should be 
satisfied by junction arrays in order to operate as quantum logic 
gates. The first set of conditions is given by the following 
string of inequalities: 
\begin{equation}
T \ll E_J \ll E_C \ll \Delta \, .  
\label{10} \end{equation} 
The two limiting energy scales in this relations, temperature $T$ 
and energy gap $\Delta$, are practically constrained by the 
available refrigeration technology and superconducting materials. 
The lower limit is set by the typical electron temperature 
attainable in experiments with the 
dilution refrigerator and is on the order of 30 mK. The upper 
limit can not be much larger than the energy gap of niobium, 
or its compounds, i.e., about 20 K. The ratio of the Josephson 
coupling energy 
$E_J$ to the charging energy $E_C$ can not be varied arbitrarily 
because of the technological limitations on the critical current 
density that can be obtained while preserving the quality of the 
tunnel junction. Conditions (\ref{10}) are satisfied if we 
take, for instance, $E_J \simeq 1$ K, and $E_C \simeq 3$ K. This 
value of $E_C$ corresponds to the junction capacitance 
$C\simeq 0.5$ fF, which for a typical specific capacitance of a 
tunnel junction, 0.1 pF/$\mu$m$^2$, requires the junction area 
of about $70\times 70$ nm$^2$. With this area, the cited $E_J$ 
value corresponds to the critical current density $j_c$ about 
10 $\mu$A/$\mu$m$^2$, and the total critical current 
$I_c =2eE_J/\hbar \simeq 50$ nA. Experimentally, this value of 
$j_c$ is within the range of current densities that can be 
achieved without the degradation of the tunnel junction quality     
\cite{b20}.  

Another condition on the junction array as a CN gate is that 
the number $N$ of junctions in it is much larger than its  
screening length: 
\[ N\gg ({C}/{C^*})^{1/2}\, . \] 
Here $C^*$ is the total stray capacitance of the array islands which 
include capacitance $C_0$ to the gate electrodes and capacitance to 
the ground planes. This condition does not represent a serious 
obstacle to realization of a CN gate. Specific values of $N$ and 
$C^*$ are dictated by the convenience of fabrication of either 
longer arrays or larger capacitances to the ground. 

The most difficult is the condition that the probability $\alpha$ 
of the decoherence-induced error during one cycle of the gate 
operation is small. Estimating the period of this cycle roughly 
as $\hbar/E_J$ we obtain from eq. (\ref{9**}) that the lower 
bound on $\alpha$ is: 
\begin{equation}  
\alpha \simeq (\frac {C_0}{C})^2 \frac{e^2 \rho}{\hbar} \, . 
\label{11} \end{equation} 
The values of parameters that are typical for existing 
experiments (in which no effort was maid to decrease $\alpha$) 
are $C_0/C \simeq 0.1$, and $\rho \simeq 300$ Ohm \cite{b21}.  
(The latter value corresponds to a narrow, about $1 \mu$m, 
electrode.) In this case $\alpha \simeq 10^{-3}$. 
The error probability can be substantially reduced by making 
coupling capacitance $C_0$ smaller, and gate electrodes wider 
thus decreasing $\rho$. Although only experiments can tell what is 
the limit to decrease in decoherence rate, it is reasonable to 
expect that $\alpha$ can be reduced further by a few orders of 
magnitude to a value about $10^{-6}$. 

In summary, we proposed a new design of the controlled-NOT quantum 
logic gate based on the adiabatic level-crossing dynamics of the 
coupled q-bits. The design is suitable for implementation in 
systems of small tunnel Josephson junctions and has the advantage 
of being insensitive to spread of the junction parameters. 
The level of decoherence in the small tunnel junction systems is 
estimated and appears to be sufficiently small for medium-scale 
quantum computation.  

The author thanks J.P. Pekola and K.-A. Suominen for critical 
reading of the manuscript. This work was supported by the ONR 
grant NOOO149610623.

\figure{ Time evolution of the energy levels of a controlled-NOT 
quantum logic gate based on the adiabatic level-crossing dynamics. 
\label{f1}}

\figure{ Energy diagram of a tunnel Josephson junction in the 
Coulomb blockade regime biased with the external voltage that 
induces the charge $Q_0 \simeq e$ on the junction capacitance. 
The two states $n=0,1$ are nearly-degenerate and the junction 
behaves effectively as a macroscopic two-level system. \label{f2} }

\figure{ Schematic layout (a) and equivalent electrostatic circuit 
(b) of an array of small Josephson junctions representing one 
q-bit. (c) The controlled-NOT gate obtained by coupling of the two 
arrays.  \label{f3} }

\end{document}